\documentstyle[12pt]{article}
\begin{document}
\vskip 72pt
\centerline{\bf STUDY OF NUCLEATING EFFICIENCY OF}
\centerline{\bf SUPERHEATED DROPLETS BY NEUTRONS}
\vskip 36pt
\centerline{B.ROY, B.K. CHATTERJEE, M. DAS and S.C. ROY}
\centerline{\it Department of Physics, Bose Institute, Calcutta 
700009, India}
\vskip 36pt

Superheated droplets have  proven to be excellent detectors for neutrons 
and could be used as a  neutron  dosimeter.  To  detect  accurately  the 
volume of superheated droplets nucleated, an air displacement system has 
been  developed.  Here  the air  expelled  due  to  volume  change  upon  
nucleation displaces a  column  of  water through  a  narrow  horizontal 
glass  tube,  and,  the displacement of water is linearly related to the 
nucleated volume and has the added advantage of being leak free.

In presence of neutrons, the rate of nucleation (rate of decrease in the 
volume  of superheated  droplets) is proportional to the residual volume 
of superheated droplets and the neutron flux  ($\phi$). Hence the volume 
of accumulated vapour (or the volume  of the displaced air) is given as:
\begin{equation}
V  =  {V_0} \left( 1 \: - \: e^{-t/\tau} \right)                             
\end{equation}
where $\tau$ = M/($\phi \rho \eta N_A$ d $\sum n_i \sigma_i$), M is  the 
molecular weight and $\rho$  is  the  density of the superheated liquid, 
$N_A$  is the Avogadro number, $n_i$  is  the  abundance of the $i^{th}$ 
species of nucleon (whose neutron elastic  scattering  cross  section is 
$\sigma_i$) in the molecule, $\eta$ is the  efficiency  of nucleation of 
the droplet by a recoil nucleon, d is the  average  droplet  volume  and  
$V_0$   is  the  volume  of  vapour of the entire superheated liquid. By 
least squares fitting of the volume  of  displaced air (V) as a function 
of time (t), $V_0$  and $\tau$ are obtained. From  $\tau$ one may obtain  
$\eta$  if  the  other  parameters  are  known. Results of an experiment 
performed  with  Freon- 12 samples using an  Am-Be  neutron  source  are 
presented in Fig. 1.

\vskip 12pt

\noindent {\bf Figure Captions}

\noindent {\bf Fig.1.} Variation of volume (scaled by the mass of sample 
and  the  area  of  cross  section  of  the  tube) of displaced air as a 
function of time  for  Freon-12; solid curve is the least-squares-fit to 
experimental data.

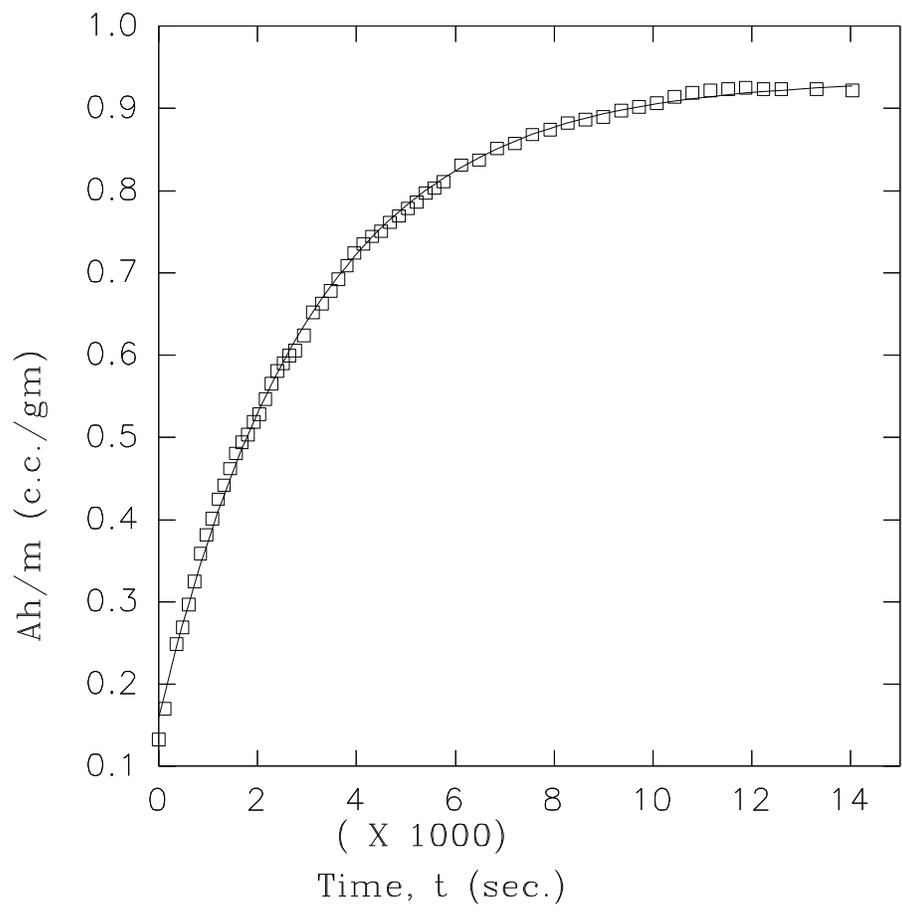
\end{document}